# From Dense Core to Halo: Amperometric Measurements and Dynamic Models Reveal a Mechanism for How Zinc Alters Neurotransmitter Release

Lin Ren,[1+] Alexander Oleinick,[2+] Irina Svir,[2] Christian Amatore[2,3]* and Andrew G. Ewing[1]*

[1] Department of Chemistry and Molecular Biology University of Gothenburg, Kemivägen 10, 41296 Gothenburg(Sweden)

[2] CNRS – École Normale Supérieure, PSL Research University – Sorbonne University, UMR 8640 "PASTEUR", Département de Chimie, 24 rue Lhomond,75005 Paris (France)

[3] State Key Laboratory of Physical Chemistry of Solid Surfaces, College of Chemistry and Chemical Engineering, Xiamen University, 361005 Xiamen, China.

[+] These authors contributed equally to this work.

E-mail: christian.amatore@ens.fr; andrew.ewing@chem.gu.se

**Abstract:** Zinc, a suspected potentiator of learning and memory, is shown to affect exocytotic release and storage in neurotransmitter-containing vesicles. Structural and size analysis of the vesicular dense core and halo using transmission electron microscopy was combined with single-cell amperometry to study the vesicle size changes induced after zinc treatment and to compare these changes to theoretical predictions based on the concept of partial release as opposed to full quantal release. This powerful combined analytical approach establishes the existence of an unsuspected strong link between vesicle structure and exocytotic dynamics which can be used to explain the mechanism of regulation of synaptic plasticity by $Zn^{2+}$ through modulation of neurotransmitter release.

Zinc is highly concentrated in neurotransmitter-containing vesicles and plays an important role in the central nervous system. Although several studies have shown that zinc is involved in learning and memory, the chemistry of how zinc changes the strength of synaptic connections and further effects the complex neuron system has been difficult to establish.[1] It has been suggested that zinc affects exocytosis, the cellular process critical for a wide range of cellular communications, especially in neurons. Micromolar zinc has been found to both reduce the amount of neurotransmitters stored in vesicles and also to slow down the dynamics of the exocytotic process, which leads to a larger fraction of release,[2] a critical component of the recently discovered partial release versus all or none exocytosis.[3] A better understanding of the pathways through which zinc changes the vesicle content and release dynamics is crucial to our understanding of its role in neurotransmission.

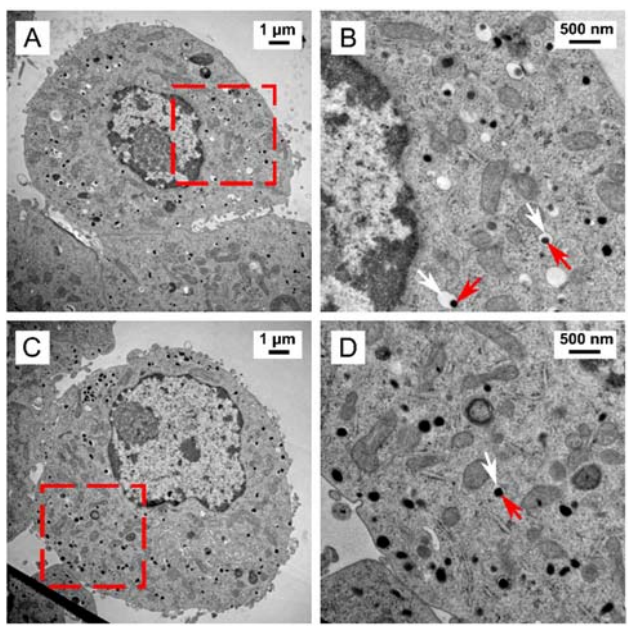

**Figure 1.** Typical TEM images for control PC12 cells (A) or after $Zn^{2+}$-treatment (C). (B) and (D) are the zooms of the areas delimited by red dashed lines in (A) and (C) respectively. Red and white arrows in (B) and (D) point to dense core and halo, respectively.

Secretory vesicles are the nanoscopic vehicles of the cell carrying soluble proteins, peptides and/or neurotransmitters. They can be divided into two types: small synaptic vesicles (SSVs) and large dense core vesicles (LDCVs). The SSVs are smaller in size (approximately 50 nm) and most common in neuronal synaptic transmission.[4] The LDCVs typically have diameters around 150−300 nm and an inner morphology defined largely by the presence of a protein core surrounded with a lucent solution called the halo (Figure 1). The dense core in LDCVs is made up of a variety of soluble materials, including proteins and nucleotides which act as a binding matrix for the transmitters. These molecular associations reduce the osmolality of the transmitter, allowing it to be accumulated to very high concentrations within the vesicle (0.1-1.0 M). Vesicle size changes in a manner that mirrors quantal size changes.[5] The structural variation of the dense core and halo reflect the difference in neurotransmitter concentrations in each compartment which can be strongly correlated to amperometric events.[6] Transmission electron microscopy (TEM) is a well-established quantitative method for measuring subcellular domains within the construct of the cell environment and, therefore, was employed for size analysis of vesicles.[7] Single-cell amperometry is the most used technique to measure exocytotic release, in which neurotransmitters excreted during exocytosis are quantified by oxidation at a carbon-fiber microelectrode placed on top of the cell membrane.[8] Hence, in this work, we carried out TEM to study the vesicle size variations of the dense core and halo with and without zinc treatment and compared these changes to theoretical predictions from amperometric current transients. With this approach we were able to establish a strong link between vesicle size and structure and exocytotic dynamics which can be used to explain the mechanism of regulation of synaptic plasticity by $Zn^{2+}$ through modulation of neurotransmitter release.

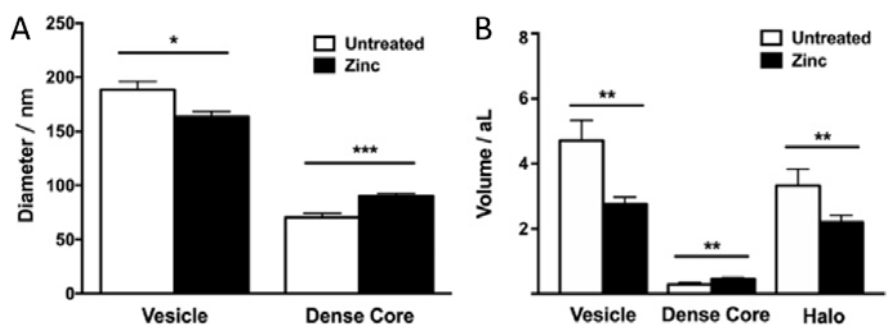

**Figure 2.** Comparison of vesicle and dense core (A) diameter and (B) volume calculated by TEM data (63 vesicles from 10 cells for control, 81 vesicles from 12 cells for zinc, only those vesicles with clear dense core and halo were chosen and calculated). Statistical significance (*: $p < 0.05$; **: $p < 0.01$; ***: $p < 0.001$).

We used PC12 (pheochromocytoma) cells, which are rich in catecholamine neurotransmitters, as a model cell line. Cells treated with or without zinc were prepared for TEM followed the steps described with detail in supporting information (SI). Representative electron micrographs for analysis of vesicles size and inner structure in single cells are pictured in Figure 1 and measurements that quantify these data are listed in Table S1. The TEM image in Figure 1A represents the prime components of the large dense core vesicles in this neurosecretory cell model, namely, dark granule composed of a highly compacted semi crystalline matrix of acidic proteins and catecholamine cations (referred to as the dense core) surrounded by a less compacted compartment of solubilized catecholamines (referred to as the halo) both being contained by the vesicle membrane as shown in the enlarged picture in Figure 1B. It is interesting to see that after zinc treatment, obviously, the halo becomes smaller compared with the control. The average diameter of vesicles measured from TEM images is $188 \pm 8$ nm (n=63; error is SEM) for control, and $164 \pm 5$ nm (n=81; error is SEM) after zinc treatment. The volumes of each intravesicular domain were then calculated based on TEM size measurements (see SI). As presented in Table S1, after zinc treatment, the volume of vesicle as well as the volume of halo decreases almost to half of control, whereas in contrast the volume of the dense core increases to almost twice that of control. More clear comparisons can be seen from Figure 2 showing bar plots of both diameter and volume measurements. All the differences are statically different with a Mann-Whitney U-test.

The morphological changes of vesicles discussed here might have consequences for exocytotic spike characteristics and transmitter release mode. Additionally, the fusion pore, the transitory connection between vesicle and cell membranes formed during exocytosis, is of significant importance. Transmitters diffuse out of the vesicle through this pore which normally closes leading to partial release or occasionally extends possibly resulting in full collapse of the secretory vesicle into the membrane.[3] In order to investigate this, representative amperometric traces were selected from the whole data set (see "Single Cell Amperometry" in SI) for complete theoretical analysis. At the beginning a sub-set of spikes with well-defined pre-spike feature (PSF; small current increase prior to the larger release event) was analyzed. The data reported in Table 1 (see also Figures S2A and S2B for graphical representation of the data) show that $Zn^{2+}$ treatment diminishes the PSF current plateau and enlarges PSF duration. Interestingly, the proportion of the spikes with PSF noticeably increases after $Zn^{2+}$ treatment (control: 36%, $Zn^{2+}$: 88%). Analysis of the PSF data provides the transmitter diffusion rate $\kappa = D/R_{ves}^2$ ($D$ is the diffusion coefficient within the polyelectrolyte matrix of radius $R_{ves}$) as discussed previously.[3d, 9, 10a,b] Comparison of $\kappa$ distribution characteristics displayed in Table 1 and Figure S2C provide evidence that the diffusion rate decreases after Zn-treatment (median values: $\kappa_{Zn}$ = 4.4×10$^3$s$^{-1}$ vs. $\kappa_{control}$ = 5.8×10$^3$ s$^{-1}$) although $(R_{ves})_{Zn}$ = 82 nm vs. $(R_{ves})_{control}$ = 94 nm as deduced from TEM (Figure 2, Table S1). Conversely, the total released charge $Q_0$ ($Q = 2Fq$, where $F$ is the Faraday constant and $q$ the number of moles of neurotransmitter) in events with PSF is not statistically different between control and Zn-treated samples (Table1 and Figure S2D). Using the above $\kappa$ median values for spikes with or without PSF allowed modelling transmitter release[3d] to extract the maximum pore sizes relative to vesicle radius ($R_{pore}^{max}/R_{ves}$) for control and Zn-treated populations (see SI for more details). Based on the mean vesicle sizes determined by TEM (Table S1) one obtains the distributions of reconstructed $R_{pore}^{max}$ values (Table 1, Figure S3). This shows that $Zn^{2+}$ treatment significantly decreases the maximum size of the fusion pore radius during exocytosis (10.7 nm) with respect to control (17.2 nm). This 60% decrease of the maximum fusion pore size is likely to result in lower diffusional flux of neurotransmitters out of the cell after Zn-treatment.

**Table 1:** Descriptive spike characteristics for control and zinc-treated PC12 cells (see text for definitions)

| Percentiles | 10% | 25% | 50% | 75% | 90% | Statistical significance [a] |
|---|---|---|---|---|---|---|
| **PSF current / pA** [b] | | | | | | |
| Control | 1.2 | 1.4 | 1.9 | 2.7 | 3.6 | * p = 0.019 |
| Zn treatment | 1.0 | 1.1 | 1.5 | 2.2 | 3.0 | |
| **PSF duration / ms** [b] | | | | | | |
| Control | 0.7 | 1.0 | 1.4 | 2.9 | 4.9 | ** p = 0.0034 |
| Zn treatment | 1.1 | 1.6 | 2.3 | 3.4 | 6.4 | |
| $\kappa$ / $10^3$ s$^{-1}$ [b] | | | | | | |
| Control | 3.1 | 4.5 | 5.8 | 7.9 | 10.3 | ** p = 0.0012 |
| Zn treatment | 2.3 | 3.1 | 4.4 | 6.4 | 8.4 | |
| $Q_0$ / fC [b, c] | | | | | | |
| Control | 15.3 | 19.5 | 33.8 | 45.9 | 54.2 | |
| | *(11.1)* | *(16.5)* | *(25.0)* | *(37.8)* | *(51.4)* | n.s. p = 0.81 (** |
| Zn treatment | 15.5 | 18.8 | 28.8 | 49.4 | 74.3 | *p = 0.0013*) |
| | *(16.3)* | *(21.4)* | *(32.2)* | *(46.8)* | *(77.9)* | |
| $R_{pore}^{max}$ / nm [d] | | | | | | |
| Control | 6.4 | 12.1 | 17.2 | 29.1 | 36.0 | *** p = 1.1·10$^{-4}$ |
| Zn treatment | 5.1 | 7.7 | 10.7 | 18.0 | 31.4 | |

[a] Statistical significance of the difference between control and Zn-treated samples was evaluated with Mann-Whitney test; usual thresholds were used to infer significance (*: p < 0.05; **: p < 0.01; ***: p < 0.001; n.s.: not significant, p > 0.05); [b] analysis performed on the set of spikes possessing well-defined PSF (pre-spike feature). Control: 59 spikes; Zn$^{2+}$: 52 spikes; [c] $Q_0$ values for all spikes (i.e., with and without PSF) and the corresponding statistical significance are indicated in italics between parenthesis, see Figure S1; [d] analysis performed on overall population of the spikes, i.e. with and without PSF. Control: 162 spikes; Zn$^{2+}$: 88 spikes.

Except for the removal of large amplitude currents ($i_{peak} > 45$ pA, Figure S4A) the Zn-treatment does not significantly affect the spike current amplitude. However, the spikes last longer (Figures S4B) leading either to same amount of released messenger molecules for spikes with PSF (Table 1) or larger amounts when all spikes are considered (+29% for the median; see Table 1 and Figure S1). These changes in spike shape are coherent with smaller time constants of the first and second exponential spike tails (Figure S5, see Experimental Section in SI for slope analysis). The augmentation of the released content after $Zn^{2+}$ exposure can be partially attributed to the increased proportion of the spikes with two-exponential tail (Figures S4C and S6). The dynamics of the fusion pore can be related to the speed of transmitter release.[3d] Evidently the smaller size of the fusion pore delays the moment where vesicle and plasma membrane merge completely and delays the release of material from large dense-core secretory granules.

As recalled above (Figure 1) PC12 cell dense-core vesicles have a dual structure.[10, 11] Catecholamine cations stored in the less compacted compartment are able to diffuse fast through the initial fusion pore.[10a,b] Conversely, those stored in the tightly packed domains (dense core) may be slowly released only after they migrate into the fast-releasing pool (halo), i.e., only after this latter one is almost emptied. Our data again support the concept of fast and slow releasing pools in the vesicle. Within this framework, release can be modeled by the following scheme:[10]

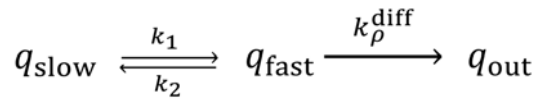

**Scheme 1**. Kinetic scheme of release within the two-pool model general framework.[10a,b]

where $q_{slow}$ and $q_{fast}$ represent the time-dependent quantities of catecholamine cations stored in the slow-releasing and fast-releasing compartments of the matrix while $q_{out}$ is the quantity that has been released. $k_1$ and $k_2$ are kinetic parameters equivalent to rate constants that feature the exchange between the two compartments (see the two-exponential regression in Fig. S6D). $k_{diff}^\rho(t) = \kappa\, R_{pore}(t)/R_{ves}$ is the pseudo-rate constant of release, incorporating geometrical and physicochemical features of a given exocytotic event.

After incubation with zinc, the vesicular volume decreases. TEM images show that the dense core is expanded while the overall vesicle is smaller, leading to a decrease of halo space and an increase in dense core/vesicle volume ratio (Figure 3A). The ensuing significant decrease in vesicle content has previously been observed by vesicle impact intracellular electrochemical cytometry after zinc treatment.[2a] This suggested a mechanism whereby zinc reversibly binds to the vesicular protein, VMAT (Vesicular monoamine transporter), leading to depletion of catecholamines from vesicles. Here, however, the expansion of the dense core and shrinking of the halo (Figure 2) indicate that the number of neurotransmitters stored in the fast-releasing pool has been decreased but those in the slow-releasing pool have increased.

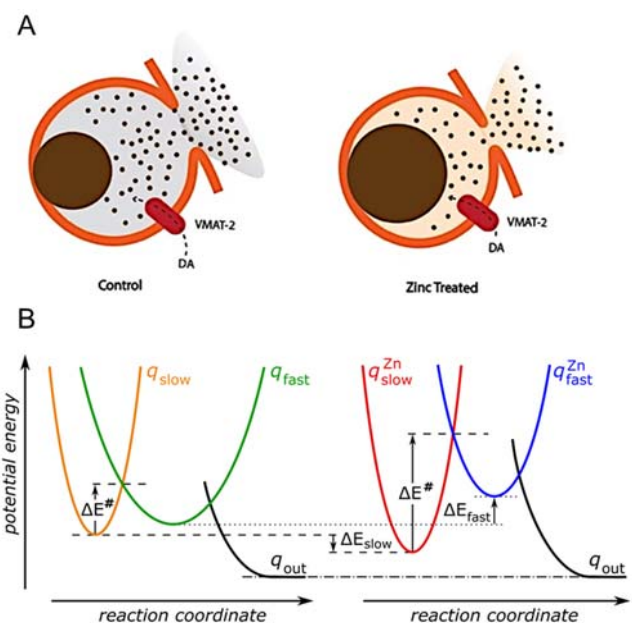

**Figure 3.** (A) Graphical representation of the changes induced by $Zn^{2+}$-treatment in releasing vesicles within the framework of the two-pool model (Scheme 1). (B) Corresponding Marcus-type representation of the $Zn^{2+}$-induced relative changes of each potential energy well featuring the neurotransmitter stability in each compartment: $Zn^{2+}$ increases its stability in the slow-releasing pool ($\Delta E_{slow} < 0$) but decreases it in the fast-releasing one ($\Delta E_{fast} > 0$) while tightening the corresponding potential energy well.

The structural changes in the vesicle after $Zn^{2+}$ treatment are consistent with a larger polyelectrolyte matrix compaction inside the vesicle upon increasing $Zn^{2+}$ concentration.[10] The simultaneous decrease of the diffusion rate in the fast pool ($\kappa$ in Table 1) also supports the view that the fast pool becomes more tightly compacted and stores fewer rapidly releasable transmitters in the presence of $Zn^{2+}$ than for controls. Therefore, one may infer that the exchange of chemical messenger between the slow and fast pools is more difficult after $Zn^{2+}$-treatment. This becomes clear through using Marcus-type diagrams of potential energies of catecholamines in each compartment for controls and for $Zn^{2+}$-treated cells (Figure 3B). The increased activation barrier, $\Delta E^{\#}$, governing the transfer of catecholamines from the slow compartment to the fast one for $Zn^{2+}$-treated cells implies that the rate constant $k_1$ (Scheme 1) is smaller for $Zn^{2+}$ than for controls. This is in full agreement with what is observed experimentally (Figure S5).

Partial release during exocytosis is a recent addition to the chemical transmission hypothesis[3] providing a mechanism for plasticity at the level of single synapses and events.[11] Amperometric analysis shows that after zinc treatment, the fusion pore stays open longer, leading to more vesicular transmitter released via the pore despite reduced diffusion rates. Mass spectrometry imaging was previously used to investigate lipid composition changes following zinc exposure.[2b] After zinc treatment, the distribution of lipid components of the cell membrane is altered and the fragments of lamellar-shaped lipids are significantly increased, whereas the fragments from high curvature lipids are reduced in the cell membrane. These lipid rearrangements within the membrane bilayers are hypothesized to contribute to fusion pore stabilization and result in changes in the pore size. These changes result in the fraction released during exocytosis increasing to more than 90%. This provides a mechanism by which zinc regulates the strength of synaptic transmission via the fraction of transmitter released. The amount of transmitter that is discharged (fraction released) into the synaptic cleft during quantal synaptic transmission depends on the product of two factors: the duration of the fusion-pore open time and the rate of neurotransmitter release.[12] When the diffusion of chemical messenger is fast, the fusion pore open-and-close time (usually short) is then unlikely to affect the amount of released transmitter. By contrast, when the diffusion of chemical messenger through the pore is slow, the prolonged release is more sensitive to the duration of the fusion pore opening and closing and might result in a larger fraction of the messenger released. This ties the regulation of the fraction of release, to changes in synaptic strength or plasticity. Indeed, assuming the same effect of zinc on synaptic vesicles, prolonged neurotransmitter release may augment recruiting of the post-synaptic receptors directly involved in long-term potentiation and hence synaptic plasticity.

In summary, we have applied single cell amperometry and TEM to investigate the effect of zinc treatment on transmitter storage and exocytotic release. The amperometric traces with and without pre-spike feature have been modelled theoretically to provide precise information on the fusion pore maximum size and kinetics of release. The vesicle size has also been studied with TEM providing evidence that changes in vesicle substructure are related to the kinetics of the exocytotic process. The finding that zinc changes intravesicular structure, fusion pore, and the rate of neurotransmitter release and therefore increases the fraction of release provides a chemical mechanism whereby zinc regulates presynaptic plasticity by increasing the fraction of chemical transmitter released to increase synaptic strength. This mechanism requires the restructuring of the vesicle chemistry to allow release during the longer open phase of fusion pore, although this reaches smaller final diameter.

**Acknowledgements**

The Gothenburg team acknowledges support from the European Research Council (ERC Advanced Grant), the Knut and Alice Wallenberg Foundation in Sweden, the Swedish Research Council (VR), and the U.S. National Institutes of Health (NIH). In Paris this work was supported by CNRS, Ecole Normale Superieure – PSL Research University, Sorbonne University (UMR 8640 PASTEUR), and the CNRS LIA 'NanoBioCatEchem'. CA acknowledges Xiamen University for his position of Distinguished Professor.